\documentclass[prb,twocolumn,preprintnumbers,superscriptaddress,amsmath,amssymb]{revtex4}

\usepackage{graphicx}% Include figure files
\usepackage{dcolumn}% Align table columns on decimal point
\usepackage{bm}% bold math

\begin{document}

%\preprint{APS/123-QED}

\title{Magnetotransport effects of ultrathin Ni$_{80}$Fe$_{20}$ films probed in-situ}

\author{Stephen Krzyk}
\affiliation{Fachbereich Physik, Universit{\"a}t Konstanz,
Universit{\"a}tsstr. 10, D-78457 Konstanz, Germany}

\author{Alexander von Schmidsfeld}
\affiliation{Fachbereich Physik, Universit{\"a}t Konstanz,
Universit{\"a}tsstr. 10, D-78457 Konstanz, Germany}

\author{Mathias Kl{\"a}ui}
\email{mathias.klaeui@uni-konstanz.de} \affiliation{Fachbereich
Physik, Universit{\"a}t Konstanz, Universit{\"a}tsstr. 10, D-78457
Konstanz, Germany}

\author{Ulrich R{\"u}diger}
\affiliation{Fachbereich Physik, Universit{\"a}t Konstanz,
Universit{\"a}tsstr. 10, D-78457 Konstanz, Germany}

%\date{\today}
\begin{abstract}

We have investigated the magnetoresistance of Permalloy (Ni$_{80}$Fe$_{20}$) films with thicknesses ranging from a single monolayer to 12~nm, grown on Al$_2$O$_3$, MgO and SiO$_2$ substrates.
Growth and transport measurements were carried out under cryogenic conditions in UHV.
Applying in-plane magnetic vector fields up to 100~mT, the magnetotransport properties are ascertained during growth.
With increasing thickness the films exhibit a gradual transition from tunneling magnetoresistance to anisotropic magnetoresistance.
This corresponds to the evolution of the film structure from separated small islands to a network of interconnected grains as well as the transition from superparamagnetic to ferromagnetic behavior of the film.
Using an analysis based on a theoretical model of the island growth, we find that the observed evolution of the magnetoresistance in the tunneling regime originates from the changes in the island size distribution during growth.
Depending on the substrate material, significant differences in the magnetoresistance response in the transition regime between tunneling magnetoresistance and anisotropic magnetoresistance were found.
We attribute this to an increasingly pronounced island growth and slower percolation process of Permalloy when comparing growth on SiO$_2$, MgO and Al$_2$O$_3$ substrates. 
The different growth characteristics result in a markedly earlier onset of both tunneling magnetoresistance and anisotropic magnetoresistance for SiO$_2$. 
For Al$_2$O$_3$ in particular the growth mode results in a structure of the film containing two different contributions to the ferromagnetism which lead to two distinct coercive fields in the high thickness regime.

\end{abstract}

\maketitle

%introduction
Spin dependent transport phenomena have become a focus of research recently with many different magnetoresistive effects (MR) being investigated.
These effects are interesting from the point of view of fundamental physics as well as for possible applications in sensors, storage and logic devices. 
In addition to the well-known anisotropic magnetoresistance (AMR) \cite{Thomson1857}, a variety of novel effects have been discovered in nanoscale sized systems, such as giant magnetoresistance (GMR) \cite{Baibich88,Binasch89} and tunneling magnetoresistance (TMR).\cite{Julliere75,Moodera95}
These effects in particular have already entered into use in industrial devices such as hard drive read heads.

Discontinuous films of ferromagnetic metals on insulating substrates can exhibit both AMR and TMR effects.\cite{Guertler01}
When increasing the thickness of the film it will start to coalesce at some point, a process which is accompanied by changes in its electrical as well as magnetic properties, and accordingly changes of the MR can be expected.  
This radical change in the transport behavior from tunneling to diffusive will also entail a radical change in the prevailing magnetoresistive effects.

%TMR
The first MR effect that will contribute to the magnetotransport is tunneling magnetoresistance, which occurs in a ferromagnet-insulator-ferromagnet (FM/I/FM) junction when the relative orientation of the magnetization of the FM contacts changes due to an external magnetic field.\cite{Julliere75}  
This field tends to align the magnetizations parallel, thus enhancing the probability for spin-dependent tunneling and lowering the resistance during application of the field.
Such a situation arises if separated islands are grown on an insulating substrate.
The probable conduction mechanism for islands with a size of tens of nanometers and separations below 10 nm is thermally activated tunneling, either substrate-assisted or through vacuum.
The tunneling conductivity of these processes drops exponentially with decreasing island distance.\cite{Neugebauer62} 
A film of separated islands consists of a large number of FM/I/FM junctions that form a complex conduction network.  
It has been shown that the magnetoresistance of such a network of ferromagnetic grains can be approximated by the magnetoresistance of a linear chain of contacts and even a single FM/I/FM junction.\cite{Vilchik06} 
Therefore the behavior of the film can be understood if we consider in the following just the growth of a few neighboring islands:

%SPM_vs_SFM_vs_FM
The magnetic behavior of a discontinuous film depends on the interplay between exchange coupling giving rise to ferromagnetism and thermal excitation leading to superparamagnetism, with a strong dependence on the size of the magnetic islands.
Using a value of $10^3 \frac{J}{m^3}$ for the magnetocrystalline anisotropy $K$ of bulk Ni$_{80}$Fe$_{20}$ \cite{Yin06} and 150 K for the film temperature $T$, we get an upper limit for the radius $r$ of a spherical superparamagnetic grain of: \cite{Handley00}

\begin{equation}
r \approx   \left( \frac{{6 k_{B} T}}{K} \right) ^\frac{1}{3}  \approx  \mbox{ 25 nm}
\end{equation}   

where $k_B$ is the Boltzman constant.
If this value is larger than the typical island size in a film prior to percolation, we can assume that the film will be superparamagnetic.
It has been shown that the magnetoresistance of a discontinuous film in this superparamagnetic regime can be modeled reasonably well by assuming just two different grain sizes. \cite{Honda97}   
The magnetization for a single grain size as a function of the external field H is then given by 

\begin{equation}
M(H) =  M_S \left( \coth (\frac{\mu H} {k{_B}T})  - \frac{k{_B}T} {\mu H} \right)
\label{Langevin}
\end{equation}  

where $M_S$ is the saturation magnetization of the film, $\mu$ the magnetic moment of each grain and T the temperature.
The tunneling probability depends on the relative orientation of the neighboring grains, leading to a dependence on the square of the total magnetization.
In the simplest approach one assumes two discrete grain sizes and then the following total magnetoresistance results:\cite{Honda97}
  
\begin{eqnarray}
\Delta R/R =   -A_S  \left( \coth (\frac{\mu_{S} H} {k{_B}T})  - \frac{k{_B}T} {\mu_{S} H} \right)^2  \\ {}-A_L  \left( \coth (\frac{\mu_{L} H} {k{_B}T})  - \frac{k{_B}T} {\mu_{L} H} \right)^2 \nonumber
\label{MR_calc}
\end{eqnarray}  

with a magnetoresistance amplitude at saturation of $A_i$ for each grain size, where the indices $S$ and $L$ are denoting the smaller and larger grain size respectively, which can be fitted.

After growing above the superparamagnetic threshold, the formation of a classic ferromagnetic phase is expected, but there are several processes which can lead to a deviation from the magnetic properties of a bulk ferromagnet.
As neighboring islands approach each other, they can couple ferromagnetically even if they are electrically separated (superferromagnetism), either via dipolar interaction \cite{Beleggia05} or tunneling exchange coupling.\cite{Bakuzis05} 
When two roughly circular islands grow together, they form a larger particle with a considerably higher aspect ratio.
This gives rise to a non-negligible shape anisotropy and increases the coercivity significantly in comparison to the value of $\leq$~1~mT for a homogeneous Ni$_{80}$Fe$_{20}$ thin film.\cite{Coburn03,Michelini02}    
Due to the statistical distribution of island sizes and separations, these magnetic phases will not occur exclusively in a certain thickness range, but rather show a gradual transition with increasing thickness.   

%AMR
As the film thickness grows, bulk MR effects set in, with anisotropic magnetoresistance (AMR) being the most important.
AMR in transition metals such as Fe and Ni is believed to be a consequence of changes in the scattering probabilities of conduction electrons due to spin-orbit interaction that depends on the relative orientation of the local magnetization and the current.\cite{Potter74} 
Phenomenologically, the AMR is characterized by a $cos^2\phi$ dependence of the resistance on the angle $\phi$ between magnetization and current direction.
For a discontinuous film it is necessary to consider that AMR only occurs intra-island, and so the observed resistance change is smaller than the intrinsic AMR effect within the islands.
For a total resistance dominated by tunneling and interface scattering, as is the case before a considerable coalescence takes place, the AMR contribution is thus expected to be negligible, but it will increase in importance with increasing film coalescence.

%motivation
Previously experiments have been carried out that have focused either on the low thickness regime where TMR dominates, or continuous films with a dominating AMR contribution,\cite{Gittleman74,Honda97,Frydman00,Guertler01,Brucas07} without studying the transition regime in detail.
Results include measurements of the TMR in a granular Ni$_{81}$Fe$_{19}$/Al$_{2}$O$_{3}$ film for various Permalloy concentrations,\cite{Brucas07} of the TMR in Ni on SiO$_{2}$ intermittently grown up to the percolation threshold,\cite{Frydman00} and
of the temperature dependence of TMR and AMR in Ni films grown on GaAs at a fixed film thickness.\cite{Guertler01}  
From these measurements it could be established, that the TMR in granular films diminishes when approaching the percolation threshold, and that there is the possibility of the coexistence of TMR and AMR in granular magnetic films.  
But so far, no study has been made available that investigated the particularly exciting thickness regime where percolation occurs, and the development of TMR as well as AMR associated with the change in the transport regime from tunneling to diffusive. 
Only with this information, a detailed understanding of the changes of the resistance and magnetic properties during this transition can be obtained. 

In this paper, we use a unique combination of in-plane magnetic vector fields and in-situ transport measurements during deposition via thermal evaporation under ultra-high vacuum (UHV) conditions to reveal the transition of the magnetoresistance response from TMR to AMR.
This measurement procedure allows us to determine the TMR and AMR for an identical film area as a function of Ni$_{80}$Fe$_{20}$ thickness without externally influencing its mechanical, electrical, thermal or chemical properties, as would be inevitable using conventional setups involving sample transfer, either in-situ or extra-situm.
From the measured magnetic and electric information we deduce the correlation between the structural growth and the magnetic properties.
For growth on different substrates, we show that magnetotransport measurements are a useful tool to identify growth conditions without the need of complicated structural in-situ scanning probe techniques.
\\
\\
%experimental
\begin{figure}[bp]
\includegraphics[width=4cm]{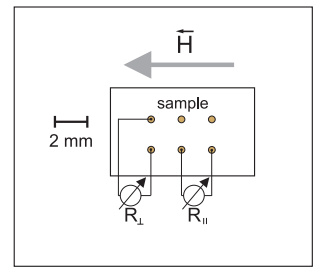}
\caption{Sketch of measurement layout. R$_{\perp}$ and R$_{\parallel}$ indicate the resistance measurement configuration for measurements with the current direction perpendicular and parallel to the magnetic field, respectively.}  
\label{layout}
\end{figure}
The samples used in this work consist of Permalloy (Ni$_{80}$Fe$_{20}$) deposited on different substrates via thermal evaporation from a rod of 6 mm diameter under UHV conditions. 
The chamber pressure during deposition was 5 $\times$ 10$^{-10}$ mbar.
Prior to insertion into UHV, the samples were cleaned with acetone and isopropanol and Au contacts were defined on the surface in a rectangular pattern with 2 mm separation.
The samples were degassed for 16 hours at 350 K and then cooled down with LN$_2$.
Permalloy was evaporated at a constant rate, varying for different samples between 0.7 and 5 nm per hour, while the film thickness is monitored using a quartz microbalance.
All thickness values mentioned in this work refer to the nominal thickness as indicated by the quartz microbalance, which corresponds to the average thickness in the surface area between two Au contacts.  
The sample resistance was measured with a commercial multimeter (Keithley 6430) in a {2-terminal} setup, covering the range from 20~M$\Omega$ to 100~$\Omega$ without changes in the measurement setup.
The measurement layout is shown in Fig. \ref{layout}. 
The in-situ vector magnet allows for the application of in-plane fields up to 100~mT at the sample position in an arbitrary orientation.
Two measurement modes were used, either field ramps from -100~mT to +100~mT and back at a constant field angle (field sweep), or rotation of the field with a constant amplitude of 20~mT (angle sweep). 
We define the value of the magnetoresistance as MR=(R(100~mT)~-~R(0~mT))/R(0~mT).
For each measurement step, several sweeps were carried out and averaged for drift and noise reduction.   
Magnetoresistance measurements were carried out at fixed film thicknesses by interrupting the deposition via a mechanical shutter and acquiring field sweep and angle sweep curves for several different contact pairs. 
After deposition of additional material, the measurements were repeated with increasing thickness up to a final thickness between 3.5 and 12~nm.
Analysis of the sample topography was carried out extra-situm by atomic force microscopy (AFM) at room temperature. 
Multiple samples with Al$_{2}$O$_{3}$, MgO  and SiO$_{2}$ substrates were investigated.
The measurements carried out show a qualitatively similar behavior for samples of the same substrate material.
\\
\\
%results+discussion
%Resistance
\begin{figure}[htbp]
\includegraphics[width=8.2cm]{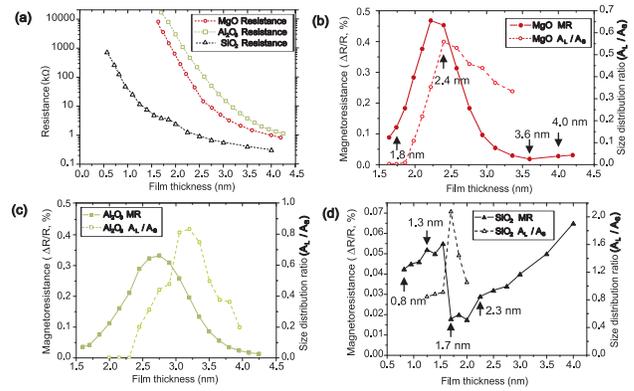}
\caption{(Color online). (a) Film resistance as function of nominal film thickness for MgO (red empty circles), Al$_2$O$_3$ (green empty squares) and SiO$_2$ (black empty triangles) samples. (b-d) The magnetoresistance (solid symbols) is shown together with the island size distribution ratio as calculated from the MR loops ($A_L/A_S$) (empty symbols) versus nominal film thickness: MgO (b), Al$_2$O$_3$ (c) and for SiO$_2$ (d) substrates. The arrows indicate the thicknesses corresponding to the curves shown in Fig.~\ref{field_sweeps_MgO}~(a-d). The resistance measurements were carried out with the current direction perpendicular to the magnetic field. The magnetoresistance values below 2.0 nm for (d) were deduced by combining the results of parallel measurements with angle sweeps. The lines are guides to the eye.}
\label{R_vs_MR}
\end{figure}
First we look at the resistance as a function of deposited Permalloy thickness (Fig. \ref{R_vs_MR}~(a)) for MgO, Al$_{2}$O$_{3}$ and SiO$_{2}$ substrates.
For all substrates the resistance initially drops exponentially with increasing film thickness, but the decrease in resistance slows significantly above some 3~nm for MgO (empty red circles) and Al$_{2}$O$_{3}$ (empty green squares) and above 1.5 nm for SiO$_{2}$ (empty black triangles).
In comparison to MgO and Al$_{2}$O$_{3}$, the resistance at a given film thickness is orders of a magnitude lower for SiO$_2$, and the difference is decreasing with increasing film thickness.

The exponential decrease at low thickness is explained by the reduction of island separation with continuing deposition.
As the film develops interconnects between islands and eventually becomes continuous, the resistance curve flattens off with higher film thickness. 
The conduction in the percolating regime is characterized by a combination of bulk-like behavior, where the resistance is caused by scattering at the lattice, impurities and defects,\cite{Ziman62} and scattering at interfaces typical for thin films.\cite{Sondheimer52} 
These effects limit the conductivity compared to the ideal ohmic resistor.
Therefore the resistance drops more slowly than the 1/d ratio expected for purely ohmic behavior.     
The markedly lower resistance for SiO$_{2}$ can be explained by a lower average distance and height of the islands for the SiO$_{2}$ sample, which leads to percolation at significantly lower nominal film thickness.
For higher thickness, all substrate types approach the continuous film state, resulting in a reduction of the conductance differences for a given film thickness. 
For low film thickness we find that the resistance decreases with increasing temperature, whereas for higher thicknesses it increases with increasing temperature. 
This is in agreement with a changeover from conduction due to thermally activated tunneling to a metallic behavior as thicker films are deposited.

\begin{figure}[htbp]
\includegraphics[width=8.2cm]{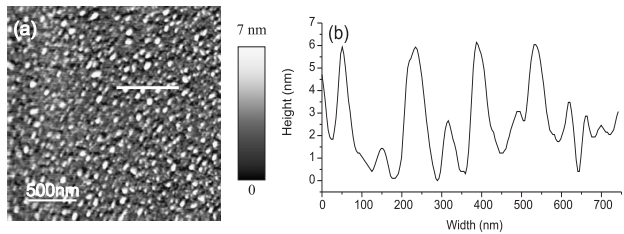}
\caption{(a) Typical Room temperature AFM image of 6~nm Permalloy evaporated at 80~K on MgO. (b)~Height profile line scan of the 6~nm Permalloy film along the trace shown by the line in (a).}
\label{AFM}
\end{figure}

\begin{figure}[bp]
\includegraphics[width=8.2cm]{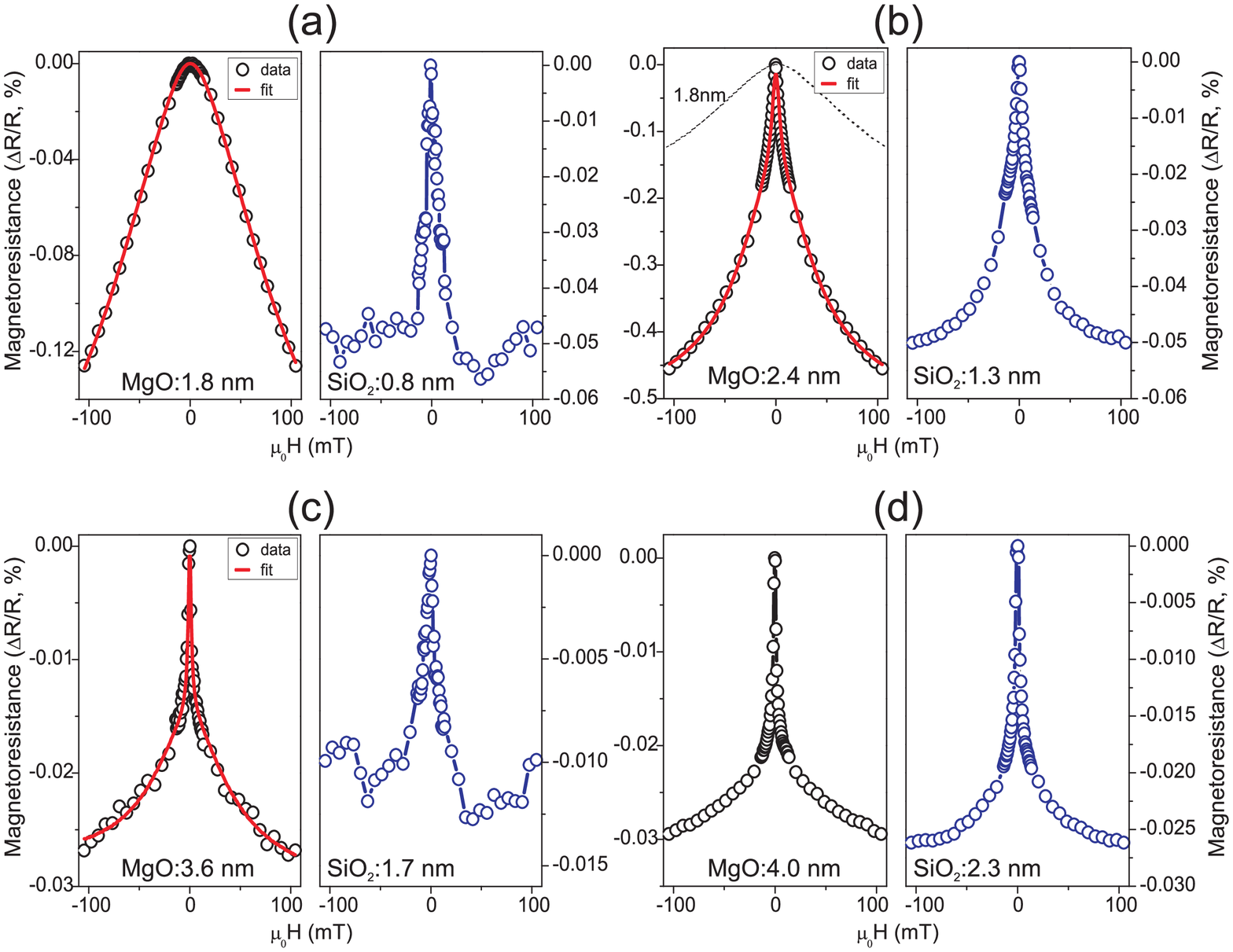}
\caption{(Color online). Field sweep MR curves of Permalloy on MgO (black circles) and SiO$_2$ (blue circles) for different film thicknesses as indicated in Fig.~\ref{R_vs_MR}~(b) and (d).
Comparable states of percolation for MgO and SiO$_2$ respectively are shown adjacent to each other: (a)~1.8~nm~MgO, 0.8~nm~SiO$_2$. (b)~2.4~nm~ MgO, 1.3~nm~SiO$_2$. (c)~3.6~nm~MgO, 1.7~nm~SiO$_2$. (d)~4.0~nm~MgO, 2.3~nm~SiO$_2$. The dotted black line in (b) shows the curve from (a) for comparison. For the MgO samples, the red lines in (a),(b), and (c) show a fit of the experimental data using the model of two different superparamagnetic island sizes (equation \ref{MR_calc}). The black and blue lines are guides to the eye. We note that we do not reach saturation of our samples for all measurements, as can be seen by the significant slope of the curves at the maximum fields for MgO.}
\label{field_sweeps_MgO}
\end{figure}

\begin{figure}[tp]
\includegraphics[width=8.2cm]{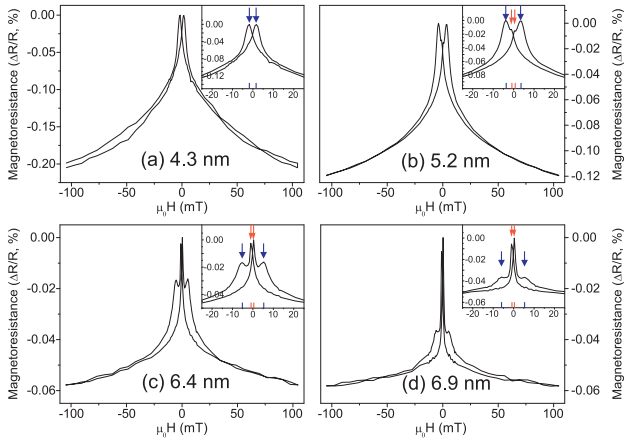}
\caption{(Color online). Field sweep MR curves of Permalloy on Al$_2$O$_3$ for film thicknesses between 4 and 7~nm. (a)~4.3~nm. (b)~5.2~nm. (c)~6.4~nm. (d)~6.9~nm. The insets show the low field regime of the field sweeps. The multiple peaks of the MR curves are indicated by arrows above the peaks and lines on the field axis (dark blue for TMR at large fields, light red for AMR at small fields). The positions of the TMR peaks move to higher field with increasing thickness, from 2~mT at 4.3~nm~(a) up to 6~mT at 6.9~nm~(d). The resistance measurements were carried out with the current direction perpendicular to the magnetic field.}
\label{field_sweeps_Al2O3}
\end{figure}

%AFM
After the end of the Permalloy deposition, extra-situm AFM images of the samples were acquired. 
For a thickness of several nanometers, the films consist of islands of roughly circular shape with a typical diameter of 50~nm (Fig. \ref{AFM} (a)).
Although single islands are clearly visible, partial overlap is evident from the height profile (Fig. \ref{AFM} (b)). 
The profile also shows an average island height of 4~nm for a film of 6 nm nominal thickness, suggesting a coalescent film.
This indicates an island growth mode on our samples, which leads to the development of separated grains for the first few nanometers of film thickness.
With additional deposition of material, these grains increase in volume, corresponding to a decrease in the average gap between grains, which eventually leads to inter-island connections and the formation of a continuous film.
\\
\\
%Magnetoresistance
Next we investigate the magnetotransport: The magnetoresistance curves of the Permalloy thin films as well as the total magnetoresistance show distinctive and non-monotonous changes with increasing film thickness.
Figs.~\ref{R_vs_MR}~(b-d) show the magnetoresistance amplitudes measured on  MgO ((b), solid red circles), an Al$_2$O$_3$ ((c), solid green squares) and a SiO$_{2}$ substrate ((d), solid black triangles) as a function of film thickness.
The magnetoresistance initially increases with increasing film thickness and reaches a maximum at about 2.3~nm for MgO (Fig. \ref{R_vs_MR}~(b)) and 2.8~nm for Al$_2$O$_3$ (Fig.~\ref{R_vs_MR}~(c)), corresponding to the thickness where the resistance curve begins to deviate from the exponential drop (see (a)).
With a further increase in film thickness, the magnetoresistance decreases.
The general trend of the magnetoresistance with increasing film thickness is similar for the MgO and Al$_2$O$_3$ sample.
The MgO sample though reaches a higher magnetoresistance and the maximum is at a lower film thickness than for the Al$_2$O$_3$ sample.
This is attributed to the dependence of the growth process on the substrate material and the statistical nature of the island growth process, leading to a different distribution of the island sizes and separations for both samples and accordingly to differences in the ratio of inter- and intra-island resistance for a given film thickness.
In comparison to MgO and Al$_2$O$_3$, the magnetoresistance as a function of film thickness for SiO$_2$ (Fig. \ref{R_vs_MR}~(d)) develops markedly differently.
Below 0.8~nm the magnetoresistance is below the noise floor, which is very high for the low thickness regime as discussed later.
Between 0.8 and 1.6~nm the magnetoresistance slightly increases to 0.05 \%, and then it drops sharply to 0.01\% above 1.6~nm and increases again above 2.2~nm film thickness.
The changes in the magnetoresistance and the different behavior for SiO$_{2}$ compared to, for instance,  MgO with increasing thickness can be explained by looking at the resistance as a function of field strength (field-sweep measurements) for different film thicknesses.
The shape of the MR curve for several thicknesses for the MgO and the SiO$_{2}$ sample from Fig. \ref{R_vs_MR} (b) and (d) respectively are shown in Fig. \ref{field_sweeps_MgO}, and these thicknesses are denoted by arrows in Fig. \ref{R_vs_MR} (b) and (d).
Note that the curves in the low thickness range for Al$_2$O$_3$ and MgO have a similar shape, although Al$_2$O$_3$ exhibits a more complex behavior with increasing thickness, which will be discussed later on (Fig.~\ref{field_sweeps_Al2O3}).
At the lowest Py thickness (1.8~nm for MgO, 0.8~nm for SiO), negative MR is clearly visible~(Fig. \ref{field_sweeps_MgO}~(a)).
The MgO curve (black line) has a smooth maximum at zero field and drops with a slow decrease of the slope with higher fields.
In contrast, the  SiO$_2$ curve (blue line) shows a triangular shape typical for TMR in multilayers or granular films, with saturation reached for 100~mT.
For the maximum MR (2.4~nm for MgO, 1.3~nm for SiO), the MR curve has narrowed for MgO, showing an increasing slope with decreasing field and a sharp maximum at zero field (Fig.~\ref{field_sweeps_MgO}~(b)).
The curve for SiO$_{2}$ has an almost identical shape, only deviating at higher fields and again reaching saturation for 100~mT, which is also the case for the next two thicknesses.
After some further 50\% increase in thickness (to 3.6~nm for MgO, 1.7~nm for SiO) the MR has dropped by more than one order of a magnitude, the curve shows an almost linear decrease with increasing field and a slightly higher slope for low fields (Fig.~\ref{field_sweeps_MgO}~(c)).
At 4.0~nm (2.3~nm for SiO$_{2}$) the MR has increased again, the curve is similar to (Fig.~\ref{field_sweeps_MgO}~(c)), but with a significantly higher peak around 0 mT (Fig.~\ref{field_sweeps_MgO}~(d)).
     
The changes of the magnetoresistance with increasing film thickness for MgO can be explained by the growth of superparamagnetic islands as follows:
At low thickness, the film consists of small islands below the superparamagnetic threshold.
When applying an external magnetic field, thermal fluctuations lead to deviations of the magnetization orientation of the islands from the parallel alignment to the field.
This causes a broadening of the resistance peak at zero field and a slow increase of the film magnetization with increasing field (Fig.~\ref{field_sweeps_MgO}~(a)). 
As the average island size increases, the influence of thermal fluctuations of the magnetization diminishes.
This results in an increase of the tunneling magnetoresistance and a more pronounced resistance peak at zero field, as can be seen by comparing Fig.~\ref{field_sweeps_MgO}~(a) and (b).
Concurrently to this process is the onset of percolation above a certain thickness, reducing the contribution of tunneling to the total resistance and therefore reducing TMR.
This leads to a decrease of the total MR in Fig.~\ref{field_sweeps_MgO}~(c). 
When the coalescence of the film has sufficiently increased, bulk-like conduction begins to contribute significantly, corresponding to the onset of AMR.
If the measurement current is orientated perpendicularly to the magnetic field, the AMR leads to a reduction of the resistance with increasing field, which is indicated by the enhancement of the peak close to zero field in Fig.~\ref{field_sweeps_MgO}~(d).
The different behavior for SiO$_{2}$ in the low thickness regime can be explained by larger islands compared to MgO, leading to a lower influence of thermal excitations on the magnetization, which is more stable due to the larger volume.
For lower temperatures the curves for the MgO substrate would have a comparable shape (as shown for the Ni/SiO$_{2}$ system at 4K in Ref. 19).
The fast increase of the magnetoresistance above 2.3 nm for SiO$_{2}$ is due to a rapidly increasing AMR above the percolation threshold.
\\
\\
This interpretation of the magnetotransport data can be underpinned by theoretical calculations based on equation \ref{MR_calc}. 
This model allows us to deduce the changes of the size distribution of the islands with increasing film thickness from the shape of the MR curves as shown in Fig. \ref{field_sweeps_MgO} (red solid lines).
Using equation \ref{MR_calc} with $A_S$, $A_L$, $\mu_S$ and $\mu_L$ as free parameters, the field sweep MR curves can be fitted very well in the TMR regime, as exemplified in Fig. \ref{field_sweeps_MgO} (a) - (c).
For higher film thickness, the curves deviate from the model that describes the situation of superparamagnetic islands, which becomes less and less valid.
For low thickness, only one pair of parameters is required, suggesting a similar size of all islands contributing to the MR.   
As the values for all fitting parameters change with increasing film thickness, we use equation \ref{MR_calc} with fixed values for $\mu_S$ and $\mu_L$ to evaluate the changes in island size.
We use the size distribution ratio $A_L/A_S$ to characterize the relative abundance of large and small superparamagnetic islands.
The changes of $A_L/A_S$ with increasing film thickness for the different sample types can be seen in Fig. \ref{R_vs_MR}~(b) - (d) (empty symbols). 
For MgO and Al$_2$O$_3$, $A_L/A_S$ mirrors the changes in the total MR (solid symbols), with a small shift to higher thicknesses.
This similar behavior is at first surprising given the fact that while $A_L$ and $A_S$ are both correlated with the total MR (see eq. 3), the ratio $A_L/A_S$, which describes the relative contribution of the large and small island types to the overall resistance and can therefore be used as a measure for the average size of the superparamagnetic islands, is not necessarily directly related to the MR.
To understand how the changes of $A_L/A_S$ affect the overall magnetoresistance, one needs to take into account the fact that the films incorporate Permalloy islands ranging in size from small nonmagnetic grains and superferromagnetic islands to larger ferromagnetic regions.
Only the intermediate sized superferromagnetic islands contribute to the TMR, and accordingly only these are affecting the shape of the TMR loops and are thus modeled. 
As we do not reach saturation at 100 mT, and larger islands reach saturation at lower fields, an increase in average island size translates to an increase in the TMR amplitude, when assuming a constant value of the saturation MR.
This is reflected in the parallel rise of the MR and $A_L/A_S$ for low film thickness in \ref{R_vs_MR}~(b) and~(c). 
When the films reach the percolation threshold, the largest islands tend to merge to form large ferromagnetic continuous areas, which are outside the model, so effectively this reduces the number of large superparamagnetic islands.
This effect is counteracting the overall increase of the size of each island due to deposition, leading to a peak of the ratio $A_L/A_S$ shortly after the onset of percolation.   
Additionally, the area fraction of the superparamagnetic islands is diminishing, so transport is to an increasing extent taking place in the ferromagnetic part of the film, thereby reducing the TMR amplitude.
As a result, the overall MR starts to decrease already before $A_L/A_S$ reaches the maximum.
With continued percolation, the contribution of superparamagnetic islands to the overall conductivity is steadily decreasing, which, in addition to the decreasing $A_L/A_S$, leads to a fast drop of the MR.   
The SiO$_2$ sample exhibits a seemingly different behavior compared to MgO and Al$_2$O$_3$. Starting from an already high value, the average island size reaches a sharp maximum at the film thickness where the MR drops significantly and then starts to decrease again. This can be understood by considering that the transition to a continuous film occurs at a lower thickness and in a rather narrow thickness range compared to MgO and Al$_2$O$_3$, corresponding to a larger island size on SiO$_2$. Therefore the SiO$_2$ sample shows more abrupt changes of the average island size during percolation, in line with the more abrupt changes in the MR.
So we see that thus our theoretical analysis confirms our interpretation of the MR evolution as a function of film thickness: The rise of the MR in the low thickness regime can be attributed to a steady increase in average island size, resulting in an increase in average magnetization for a given magnetic field strength and an increase of the TMR. With the onset of percolation, the average island size is decreasing and the superparamagnetic fraction of the film is reduced, leading to a reduction of the overall MR, up to the thickness where AMR starts to contribute significantly to the MR. 
\\
\\ 
\begin{figure}[bp]
\includegraphics[width=8.2cm]{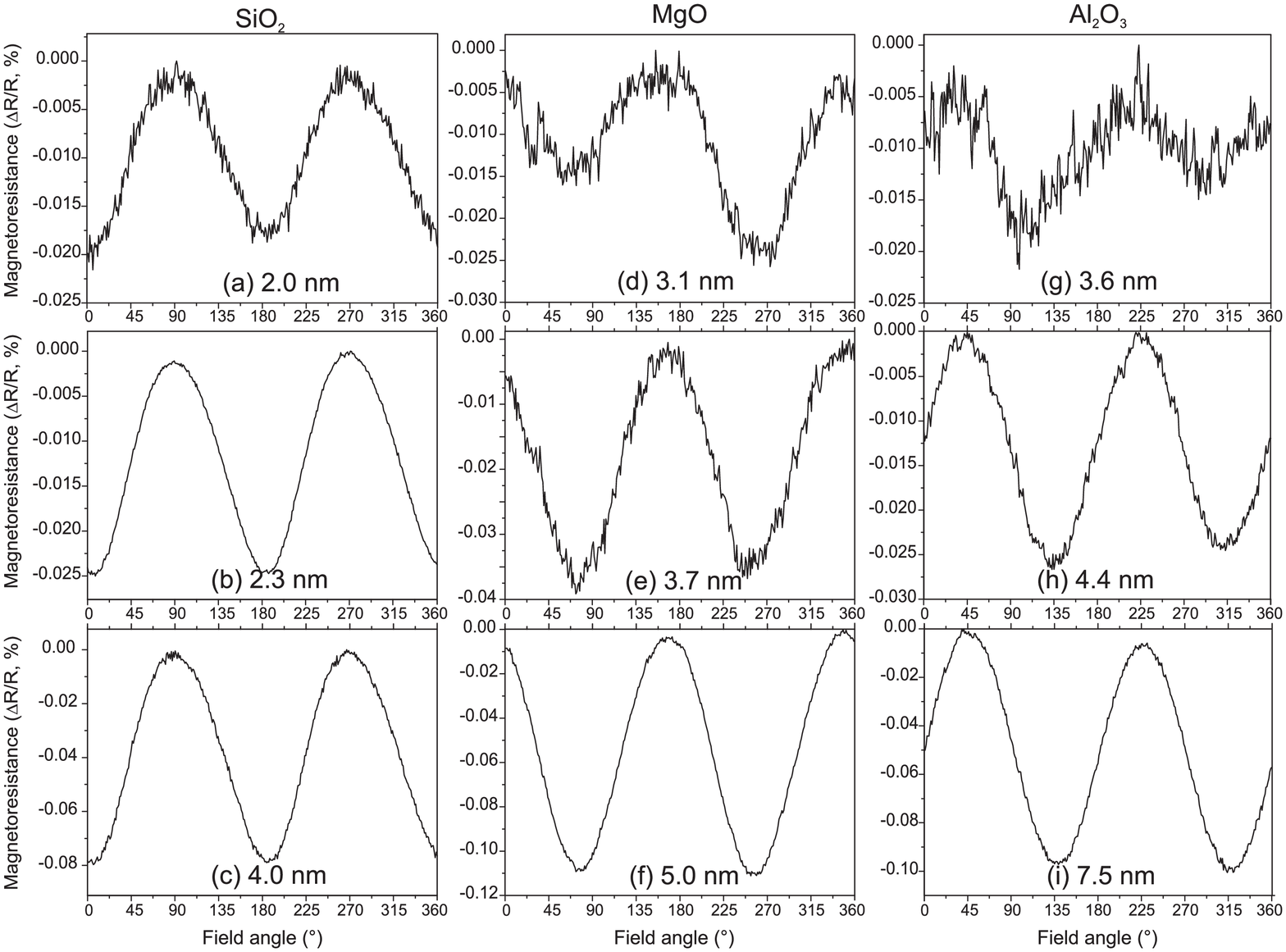}
\caption{Angle sweep MR curves of Permalloy grown on SiO$_2$, MgO and Al$_2$O$_3$ substrates for various film thicknesses up to 7.5~nm. (a)~SiO$_2$~2.0~nm. (b)~SiO$_2$~2.3~nm. (c)~SiO$_2$~4.0~nm. (d)~MgO~3.1~nm. (e)~MgO~3.7~nm. (f)~MgO~5.0~nm. (g)~Al$_2$O$_3$~3.6~nm. (h)~Al$_2$O$_3$~4.4 nm. (i)~Al$_2$O$_3$~7.5~nm. Field angles of 90$^{\circ}$, 170$^{\circ}$ and 45$^{\circ}$ correspond to the parallel orientation of magnetic field and current direction for the SiO$_2$, MgO and Al$_2$O$_3$ substrates, respectively.}
\label{angle_sweeps_Al203}
\end{figure}
As can be inferred from the shape of the MR curves in Fig.~\ref{field_sweeps_MgO}~(d), the contributions of AMR and TMR appear superimposed in the field-sweep measurements.
The magnetoresistance in the TMR thickness regime was found to be isotropic, as expected for a random distribution of island sizes and distances.
Under the assumption that this is also the case for higher film thickness, angle sweep measurements can be used to separate the contributions of the angle-dependent AMR and the isotropic TMR.
The development of the AMR for Permalloy grown on MgO, Al$_2$O$_3$ and SiO$_{2}$ substrates is shown in Figs.~\ref{angle_sweeps_Al203}~(a-c) for a SiO$_{2}$ sample, Figs.~\ref{angle_sweeps_Al203}~(d-f) for a MgO sample and in Figs.~\ref{angle_sweeps_Al203}~(g-i) for a Al$_2$O$_3$.
The onset of the AMR occurs at the lowest thickness for SiO$_2$ followed by MgO and by Al$_2$O$_3$. 
The AMR exhibits a clear $cos^2\phi$ oscillation of the resistance for all samples, and increases with increasing thickness. 
A notable feature is the significantly higher noise at the onset of AMR compared to larger thicknesses for all substrates.
This is clearly visible, when comparing the angle-sweep curves at the onset of AMR, and after a slight increase in thickness (Fig.~\ref{angle_sweeps_Al203}~(a) versus~(b)) for SiO$_2$,  Fig.~\ref{angle_sweeps_Al203}~(d) versus~(e) for MgO and Fig.~\ref{angle_sweeps_Al203}~(g) versus~(h) Al$_2$O$_3$.
Although the signal amplitudes differ by less than a factor of 2, the noise is considerably higher for the lower thicknesses. 
This is consistent over all measurements, making it impossible to detect AMR below a relative value of some $10^{-4}$.
Sousa et al. have attributed this behavior to Barkhausen jumps of domain walls in the percolating film,\cite{Sousa04} this effect could also be due to an increased sensitivity of the conductivity to random fluctuations of the atomic structure of the Permalloy film in the percolating regime.   
\\
\begin{figure}[tp]
\includegraphics[width=8.2cm]{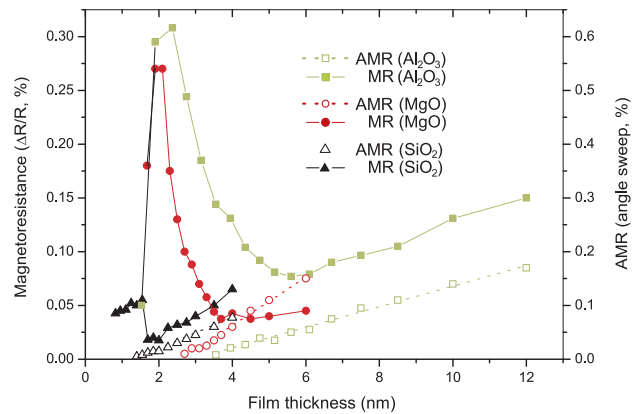}
\caption{(Color online). Magnetoresistance measured by field sweeps with the current direction perpendicular to the magnetic field (solid green squares for Al$_2$O$_3$, solid red circles for MgO and solid black triangles for SiO$_2$) and AMR as extracted from angle sweep measurements (empty green squares for Al$_2$O$_3$, empty red circles for MgO and solid black triangles for SiO$_2$) for Permalloy grown on Al$_2$O$_3$, MgO and SiO$_2$ substrates. The lines are guides to the eye.}
\label{AMR_vs_TMR_12}
\end{figure}
The increase of the AMR as a function of Permalloy film thickness is shown in Fig.~\ref{AMR_vs_TMR_12} for an Al$_2$O$_3$ substrate (empty green squares), for a SiO$_2$ substrate (empty black triangles) and MgO (empty red circles).
After its onset (for instance for Al$_2$O$_3$ between 3 and 4~nm film thickness), the AMR increases roughly linearly with increasing thickness. 
This is in good agreement with measurements in continuous multilayer films, where a linear increase of the AMR up to 10~nm Permalloy thickness was found.\cite{Thanh07}  
In comparison to the magnetoresistance values extracted from field-sweep measurements, which are also shown in Fig.~\ref{AMR_vs_TMR_12} (solid symbols), the onset of the AMR coincides with the drop of the TMR, which is for the Al$_2$O$_3$ sample followed by a transition regime up to about 7~nm thickness, where both AMR and TMR occur.
Above 7 nm only AMR contributes to the total magnetoresistance, therefore AMR and MR rise in proportion.
The MR effect for the SiO$_2$ sample shows a analogous behavior, but at significantly lower thicknesses.
Results for the MgO samples lie in between Al$_2$O$_3$ and SiO$_2$.
\\
\\
For Al$_2$O$_3$ the transition between TMR and AMR is even more directly visible in the field-sweep measurements, as shown in Fig.~\ref{field_sweeps_Al2O3}~(a-d).
At 4.3~nm thickness, only a TMR curve is observable, with peaks (dark blue arrows) at $\pm$~2~mT (Fig.~\ref{field_sweeps_Al2O3}~(a)~inset).
As the thickness is increased, AMR peaks (denoted by light red arrows in Fig.~\ref{field_sweeps_Al2O3}~(b)) appear superimposed on the TMR curve.
The TMR peaks are visible at a higher field of $\pm$~4~mT.
With further increase in film thickness the AMR peaks become more pronounced, staying at low fields close to zero.
In contrast, the TMR peak heights are reduced and they move to even higher absolute fields ($\pm$~5~mT in Fig.~\ref{field_sweeps_Al2O3}~(c)).
At 6.9~nm, the TMR is barely visible and the peaks are at about $\pm$~6~mT, while the AMR peaks at low field dominate (Fig.~\ref{field_sweeps_Al2O3}~(d)). 
For the MgO and SiO$_2$ samples on the other hand, no differences in the coercive fields detected by the TMR and AMR were observed, while the value of the coercive field remains below 1~mT (see Figs.~\ref{field_sweeps_MgO}~(b-d)).
\\
The observation of two different coercive fields for the TMR and the AMR effect for the Al$_2$O$_3$ samples (Fig.~\ref{field_sweeps_Al2O3}) indicates the presence of two different contributions to the ferromagnetism in the TMR/AMR transition regime.
These can be identified as the bulk ferromagnetism of continuous film areas which gives rise to the AMR and shows accordingly a low coercivity, and either superferromagnetism or multi-island ferromagnetism that give rise to an enhanced coercivity of the small islands that contribute to the TMR.
As these processes are very sensitive to island size, shape and separation, this implies that differences in the distribution of island size and separation exist for Al$_2$O$_3$ compared to MgO and SiO$_2$ substrates, and have a clear influence on the magnetoresistive behavior only in the percolating regime.  
A possible explanation for this surprising behavior exhibited by the Al$_2$O$_3$ substrates is indicated by the higher film thickness where percolation occurs and the broader thickness range of the transition regime from TMR to AMR for  Al$_2$O$_3$ substrates in comparison to the other substrate types.
This means, that the morphological transition from separated islands to a continuous film is rather gradual for Al$_2$O$_3$ substrates, with a broad intermediate regime where separate islands with inter-island tunneling transport exist, and at the same time continuous film areas are present where diffusive transport with AMR occurs.
In this transition regime both conduction mechanisms can contribute to a similar extent to the resistance of the film.
\\
\\     
%conclusion
In conclusion, we have determined the magnetotransport properties of ultrathin Permalloy (Ni$_{80}$Fe$_{20}$) films on insulating substrates.
The films exhibit a change from the TMR regime to the AMR regime with increasing thickness, with a transition zone where both AMR and TMR are present.
The MR loops in the TMR regime can be fitted using a theoretical model and a remarkable agreement between the MR amplitude and the island size distribution is found.
The transition to AMR corresponds to the percolation process with continuing material deposition and the transition from superparamagnetic behavior to ferromagnetism.
The film thickness dependence of the resistance, the TMR and the AMR is qualitatively similar for Al$_2$O$_3$, MgO and SiO$_2$ substrates for the thickness regimes where the MR response is clearly dominated by either TMR or AMR.
SiO$_2$ exhibits a markedly earlier onset of both tunneling magnetoresistance and anisotropic magnetoresistance, due to an onset of percolation at lower thickness.
In the transition zone from the TMR to the AMR regime we find markedly different behavior for the different substrate materials.
SiO$_2$ shows a very abrupt reduction in TMR within a thickness increase of 0.2~nm whereas for Al$_2$O$_3$ and MgO the transition is broader.

In particular for Al$_2$O$_3$ the transition zone covers a larger thickness range (3~nm) and this is accompanied by the TMR and the AMR effect revealing different coercivities.
This in turn points to the slower percolation process and a large thickness range where tunneling between islands as well as intra-island transport contribute significantly to the conduction for Permalloy films grown on Al$_2$O$_3$ in comparison to MgO and SiO$_2$.
So even though polycrystalline growth occurs on all substrates, the transition zone exhibits reproducibly very different behavior for the different substrate materials, further highlighting the importance of the substrate even if no epitaxy is used.
Our results clearly demonstrate that in-situ transport measurements are a capable tool for in-detail investigations of growth conditions of ultrathin magnetic films.

%acknowledgements
The authors acknowledge support by the DFG (SFB 513 and 767) and the Landesstiftung Baden-W{\"u}rttemberg. 

%bibliography


\begin{thebibliography}{00}
\bibitem{Thomson1857} W. Thomson, Proc. Royal Soc. \textbf{8}, 546 (1857).
\bibitem{Baibich88} M.N. Baibich \textit{et al.}, Phys. Rev. Lett. \textbf{61}, 2472 (1988).
\bibitem{Binasch89} G. Binasch, P. Gr{\"u}nberg, F. Saurenberg, and W. Zinn, Phys. Rev. B \textbf{39}, 4828 (1989). 
\bibitem{Julliere75} M. Julliere, Phys. Lett. A \textbf{54}, 225 (1975).
\bibitem{Moodera95} J.S. Moodera, L.S. Kinder, T.M. Wong, and R. Meservey, Phys. Rev. Lett. \textbf{74}, 3273 (1995).
\bibitem{Guertler01} C.M. G{\"u}rtler, Y.B. Xu, and J.A.C. Bland, J. Magn. Magn. Mater. \textbf{226}, 655 (2001).
\bibitem{Neugebauer62} C.A. Neugebauer and M.B. Webb, J. Appl. Phys. \textbf{33}, 74 (1962).
\bibitem{Vilchik06} H. Vilchik, A. Frydman and R. Berkovits, phys. stat. sol. (c) \textbf{3}, 288 (2006).
\bibitem{Yin06} L.F. Yin \textit{et al.}, Phys. Rev. Lett. \textbf{97}, 067203 (2006).
\bibitem{Handley00} R. O'Handley, \textit{Modern Magnetic Materials} (John Wiley and Sons, New York, 2000).
\bibitem{Honda97} S. Honda, T. Okada, M. Nawate, and M. Tokumoto, Phys. Rev. B \textbf{56}, 14566 (1997).
\bibitem{Beleggia05} M. Beleggia, Y. Zhu, S. Tandon and M. de Greaf, Appl. Phys. Lett. \textbf{87}, 202504 (2005).
\bibitem{Bakuzis05} A.F. Bakuzis and P.C. Morais, J. Magn. Magn. Mater. \textbf{285}, 145 (2005). 
\bibitem{Michelini02} F. Michelini \textit{et al.}, J. Appl. Phys. \textbf{92}, 7337 (2002).
\bibitem{Coburn03} R.P. Cowburn, J. Appl. Phys. \textbf{93}, 9310 (2003).
\bibitem{Potter74} R.I. Potter, Phys. Rev. B \textbf{10}, 4626 (1974).
\bibitem{Gittleman74} J.I. Gittleman, B. Abeles, and S. Bozowski, Phys. Rev. B \textbf{9}, 3891 (1974).
\bibitem{Brucas07} R. Bru\v{c}as \textit{et al.}, J. Appl. Phys. \textbf{101}, 073907 (2007).
\bibitem{Frydman00} A. Frydman, T.L. Kirk, and R.C. Dynes, Solid State Commun. \textbf{114}, 481 (2000).
\bibitem{Ziman62} J.M. Ziman, \textit{Electrons and Phonons} (Oxford University Press, London, 1962) p. 486.
\bibitem{Sondheimer52} E.H. Sondheimer, Advan. Phys. \textbf{1}, 1 (1952).
\bibitem{Sousa04} J.B. Sousa \textit{et al.}, J. Appl. Phys. \textbf{96}, 3861 (2004).
\bibitem{Thanh07} N.T. Thanh \textit{et al.}, J. Appl. Phys. \textbf{101}, 053702 (2007). 
\end{thebibliography}
\end{document}